\documentclass[12pt,a4paper]{article}
\usepackage{amsmath}
\usepackage{amssymb}
\usepackage{graphicx}
\usepackage{epsfig}
\usepackage{psfrag}

\newcommand{\av}[1]{\langle #1 \rangle}

\newcommand{\Tr}{\operatorname{Tr}}

\newcommand{\nc}{\newcommand}

\nc{\be}{\begin{equation}}
\nc{\ee}{\end{equation}}
\nc{\bea}{\begin{eqnarray}}
\nc{\eea}{\end{eqnarray}}
\nc{\wlambda}{{{\widetilde\lambda}}}
\nc{\hlambda}{{{\widehat\lambda}}}
\nc{\hh}{{{\widehat{h}}}}

\nc{\Nt}{N_{\tau}}
\nc{\Ns}{N_S}

\nc{\cA}{{\cal A}}
\nc{\cB}{{\cal B}}
\nc{\cC}{{\cal C}}
\nc{\cD}{{\cal D}}
\nc{\cE}{{\cal E}}
\nc{\cF}{{\cal F}}
\nc{\cG}{{\cal G}}
\nc{\cH}{{\cal H}}
\nc{\cI}{{\cal I}}
\nc{\cJ}{{\cal J}}
\nc{\cK}{{\cal K}}
\nc{\cL}{{\cal L}}
\nc{\cM}{{\cal M}}
\nc{\cN}{{\cal N}}
\nc{\cO}{{\cal O}}
\nc{\cP}{{\cal P}}
\nc{\cQ}{{\cal Q}}
\nc{\cR}{{\cal R}}
\nc{\cS}{{\cal S}}
\nc{\cT}{{\cal T}}
\nc{\cU}{{\cal U}}
\nc{\cV}{{\cal V}}
\nc{\cW}{{\cal W}}
\nc{\cX}{{\cal X}}
\nc{\cY}{{\cal Y}}
\nc{\cZ}{{\cal Z}}
\nc{\bk}{{{\bf k}}}
\nc{\bx}{{{\bf x}}}

\nc{\simo}[1]{{\stackrel{#1}{\simeq}}}
\nc{\geqo}[1]{{\stackrel{#1}{\geq}}}
\nc{\geo}[1]{{\stackrel{#1}{>}}}
\nc{\guo}[1]{{\stackrel{#1}{\succ}}}

\nc{\Eq}{Eq.~}
\nc{\nr}[1]{(\ref{#1})}
\newlength{\figwidth}
\begin{document}

\begin{center}

{\Large\bf Three dimensional finite temperature $SU(3)$ gauge theory
near the phase transition}
\\[1.8ex]
{\bf P. Bialas$^{a,b;}$\footnote{pbialas@th.if.uj.edu.pl},
L. Daniel$^{a}$
A. Morel$^{c;}$\footnote{andre.morel@cea.fr},
B. Petersson$^{d,e;}$\footnote{bengt@physik.hu-berlin.de}}
\\[1.2mm]
$^a$ Institute of Physics, Jagellonian University\\
ul. Reymonta 4, 30-059 Krakow Poland \\[1.2mm]
$^b$ Mark Kac Complex Systems Research Centre\\
Jagellonian University, ul. Reymonta 4, 30-059 Krakow, Poland
\\[1.2mm]
$^c$ Institut de Physique Th\'eorique de Saclay, CE-Saclay \\
F-91191 Gif-sur-Yvette Cedex, France
\\[1.2mm]
$^d$ Fakult\"at f\"ur Physik, Universit\"at Bielefeld \\
P.O.Box 10 01 31, D-33501 Bielefeld, Germany \\[1.5ex]
$^e$ Humboldt-Universit\"at zu Berlin, Institut f\"ur Physik, \\
Newtonstr. 15, D-12489 Berlin, Germany
\end{center}

\vspace{1.5cm}

\begin{abstract}

\medskip
\noindent
We have measured the correlation function of Polyakov loops 
on the lattice in three dimensional
$SU(3)$ gauge theory near its finite temperature phase transition.
Using a new and powerful application of finite size scaling,
we furthermore extend the measurements of the critical
couplings to considerably larger values of the lattice sizes, both in the temperature and
space directions, than was investigated earlier
in this theory. With the help of these measurements
we perform a detailed finite size scaling analysis, showing that for the 
critical exponents of the
two dimensional three state Potts model 
the mass and the susceptibility fall on unique scaling curves. This
strongly supports the expectation that the gauge theory is in the same universality class. 
The Nambu-Goto string model
on the other hand predicts that the exponent $\nu$ has the mean field value,
which is quite different from 
the value in the abovementioned Potts model. 
Using our values of the critical couplings we also determine the continuum limit of
the value of the critical temperature in terms of the square root of
the zero temperature string tension. This value is very 
near to the prediction of the Nambu-Goto string model
in spite of the different critical behaviour.
\end{abstract}

\section{Introduction.}

\medskip
\noindent
Three dimensional SU(3) gauge theory has many properties in common
with QCD. Lattice simulations of the theory show that the static quark potential
is linear, implying confinement. They also give evidence for 
a mass gap and a nontrivial
glueball spectrum \cite{teper}.  At finite temperature there is a
phase transition to a state in which the energy density is approximately
described by a gluon gas. It can therefore be expected that from this
model one obtains important information about the mechanism of
confinement and the deconfinement transition in QCD. Furthermore,
there exist analytic approximations
in three dimensional $SU(N)$ gauge theory 
which predict the string
tension and the glue ball spectrum at zero temperature
\cite{karabali,leigh}.
These results are in quite good
agreement with lattice calculations \cite{bringoltz,tepera}. It is
therefore important to measure the string tension at finite
temperature and the value of the critical temperature,
because the analytic calculations should eventually be
extended to these observables.

In a series of papers \cite{bialas1, bialas2, bialas3, bialas4},
we have investigated the $SU(3)$ gauge theory at finite temperature
in two spatial dimensions 
using lattice simulations. In \cite{bialas1}
we have shown that in the high temperature phase 
above approximately $1.5 T_c$, where $T_c$ is the critical temperature,
the theory can be dimensionally reduced
to a gauge-Higgs model in two dimensions, by which one can give an excellent description
of the long distance properties of the full theory. In \cite{bialas2} we have analysed the 
two dimensional gauge-Higgs model
in great detail. In \cite{bialas3} we have investigated the
thermodynamics of the three dimensional theory in the high temperature phase. We have shown, in particular, that
the trace of the energy momentum tensor has a non-perturbative behaviour in
a region above the phase transition,  analogous to the results found in
$SU(3)$ in $3+1$ dimension \cite{boyd} and in full QCD \cite{cheng, borsanyi}.
A detailed investigation of the thermodynamics of $SU(N)$ theories
in $2+1$ dimensions with $N=2$ to $6$ has more recently been performed in Refs. \cite{panero,panero2}.
 
In \cite{bialas4} we
studied the theory in the low temperature phase. By measuring the correlation
function of Polyakov loops we obtained the finite temperature string
tension. We showed that
it can be very well described by the Nambu-Goto string
model, as was predicted in \cite{pisarski, olesen}, but only up to a temperature
$T\simeq 0.7T_c$, where $T_c$ is the critical temperature.
One should remark, however,
that there are analytic calculations for a general fluctuating bosonic string, 
which in three dimensions give universal values
 for the terms in the expansion in $(T/T_c)^2$  up to $(T/T_c)^6$ \cite{luscher1,
 luscher2, aharony}.
Not only do these terms, of course, coincide with those from the development of
the formula in
Nambu-Goto string model to this order, but their contribution is practically indistinguishable 
from the full model up to $T\simeq 0.75T_c$. 
However, such a short expansion cannot give any hint about a phase transition, 
while the Nambu-Goto model predicts the existence of a critical temperature $T_c$
at which the string tension vanishes.  Moreover, it 
gives a value for the non perturbative ratio
$T_c/\sqrt{\sigma_0}$, where $\sigma_0$ is the zero temperature string tension. 
The ratio only depends on the number of transverse dimensions. 
 It does
not depend on the group $SU(N)$.  
In fact, the Nambu-Goto string model
gives a result for the approach of the finite temperature string tension to $T_c$, 
which corresponds to the mean field exponent $\nu=1/2$. 
>From universality arguments
it is, however, expected that the transition in the $SU(3)$ gauge theory 
in two spatial dimensions is in the universality class of the two dimensional
three state Potts model \cite{svetitsky}. Support for this proposal has been found
in lattice calculations \cite{legeland}. 

 Other comparisons of the Polyakov loop
correlations in SU(3) in two spatial dimensions
 with the string model, in particular at lower temperatures, 
have been performed with a different technique
in \cite{teper2}. Recently, they have been extended to
the group SU(6) \cite{teper3}. In \cite{panero} the string model is used
to estimate the behaviour of the pressure just below the transition. A good
agreement with the numerical data is found.

In this article we report on an investigation of the model 
very near to
the phase transition, a region which has not been studied up to now. 
First we calculate the critical couplings $\beta_c(\Nt)$, where
$\Nt$ is the lattice extent in the temperature direction, extending earlier
calculations \cite{legeland, liddle} to 
considerably higher statistics and larger lattices. 
For this purpose we introduce a new and powerful method 
based on the work by Binder on finite size scaling \cite{binder}.
Further, using results from the literature
for $a\sqrt{\sigma_0}$, where $a$ is the lattice spacing
 we obtain from the continuum extrapolation a value for 
$T_c/\sqrt{\sigma_0}$, which can be directly compared with the 
prediction of the string model. 

We further investigate the correlation function of the Polyakov loops below but near
the critical coupling. We extract the mass and the susceptibility for a large number
of spatial lattice sizes and for a range of couplings near the critical one. 
We consider the finite size 
scaling functions of these observables and find an impressive 
agreement with the universality class of the two dimensional
three state Potts model. However, 
although the string and the gauge 
models thus have quite different critical behaviour, 
we find that their critical temperatures are extremely close to each other.
 
The plan of the paper is as follows.
The definitions relative to the Polyakov loop correlations, and the set up
for a critical scaling analysis of the mass gap and the susceptibility
are given in Section 2. In section 3,  
$T_c/g^2$ is computed and  
the scaling properties  of the susceptibility and  mass gap as functions
of $T/T_c$ exhibited. The ratio $T_c/\sqrt{\sigma_0}$ is finally 
computed and the gauge and string models compared.  
A summary and conclusions are proposed in a last section.

\section{The Polyakov loop correlation function. Loop susceptibility and mass gap.}

A discussion of the simulations was given in \cite{bialas4}. We have
used the same algorithm, which was ported to GPUs (Graphical Processing
Units).  In this section, we first give the formulae needed for our
new analysis of the physical quantities of interest, and then recall
the scaling properties expected from universality near the transition.

\subsection{The model and the correlation function.}

The action used in the simulations is the standard Wilson lattice gauge action
\be
S_W(U_{\mu}(\bar{x},\tau))=\beta \sum_P (1-\frac{1}{3}Re\Tr U_P),
\ee
where $U_{\mu}(\bar{x},\tau)$ denotes the group element on the link in the 
direction $\mu$, whose origin is located at $\bar{x}= (x_1,x_2)$ in space and at $\tau$
in the temperature direction. 
The lattice has
extension $\Ns^2\times \Nt$.
The variables $(x_1, x_2, \tau)$ are defined
on integer values, $x_1,x_2 = 0,1, ... ,\Ns-1$ and $\tau=0,1,...,\Nt-1$.
The matrix $U_P$ is the product of link matrices around a plaquette.
The constant $\beta$ is the lattice coupling constant. Note that 
in this article we will
never use $\beta$ to denote the inverse temperature.
The temperature and volume of the lattice are defined by

\bea
\frac{1}{T} & = & aN_{\tau}, \label{temp}\\
V & = & (aN_S)^2.  \label{vol}
\eea

\noindent
where $a$ is the lattice spacing.
The coupling constant $\beta$ in the action 
is related to the coupling constant $g$ of the continuum action
by 
\be
\lim_{a\rightarrow 0}(a\beta)=\frac{6}{g^2}.  \label{g2}  
\ee
In three space time dimensions, $g^2$ has dimension of energy, and can thus
be used as the energy scale in the continuum theory. 

When $\Ns \rightarrow \infty$ 
at fixed $\Nt$ there 
is a phase transitions with a critical coupling
$\beta_c(\Nt)$ to a state where the $Z_3$ symmetry of the theory is broken.
This corresponds to a finite temperature transition into a deconfined state.

To investigate the theory around the phase transition we performed
numerical simulations in the neighbourhood of the transition 
for a large number of values of the coupling constant $\beta$ 
and lattice extensions $\Ns$ and $\Nt$. We study the order
parameter, which in this case is the Polyakov loop.
The local Polyakov loops are winding around the temperature direction:

\be
L(\bar{x}) = Tr \prod_{\tau=0}^{N_{\tau}-1}U_{\tau}(\bar{x},\tau)
\label{polyak}.
\ee

\noindent
We define the projected correlation function $G(z)$ between two loops at distance $z$
from each other 
in the $x_1$ direction by

\be
G(z) = \frac{1}{N_S}\sum_{x_2} 
Re\av{L(\bar {0})L^*(\bar {x}=(z,x_2))}.
\ee
\noindent

In our earlier work \cite{bialas4}, devoted to couplings well below 
$\beta_c(\Nt)$, we have represented this correlation function by  
\be
G(z) = b \cosh \left[ m \left( \frac{\Ns}{2} - z \right) \right].
\label{gfit}
\ee
The subtraction of a disconnected part from $G(z)$ would be required 
above $\beta_c(\Nt)$.
\noindent
In (\ref{gfit}), the coefficient $b$ and the mass $m$ were fitted
to the data at $z$ larger than some short distance cut off $z_0$. 
This allowed us to extract the mass gap $m$
and thereby the temperature dependent string tension 

\be
\frac{\sigma(T)}{T^2} = m \Nt \label{msigma}.
\ee

\noindent

Fits with Eq.(\ref{gfit}) 
must give stable results  with respect to changes of $z_0$, in which case
they directly 
provide a reliable estimate of $m$, thus of the largest 
correlation length $\xi=1/m$. 
It corresponds to the simplest
situation where the correlation in $z$ decays as a pure exponential at
large enough distances, and this was the case in the analysis which  
we reported in \cite{bialas4}.
 
When we approach the phase transition,
we find that this procedure does not work. In fact, making a fit with Eq.
(\ref{gfit}) the mass $m$ changes whatever $z_0$ we choose. Instead of trying a
complicated fit in coordinate space, we prefer to transform the results
to momentum space, where the analysis can be made more systematically.
 
The correlation function in (\ref{gfit}) 
can be as well represented in momentum space
by a simple pole, in the statistical mechanics called
the Ornstein-Zernike (OZ) behaviour. In general, the analytic behaviour
in momentum space on the negative real axis of the square of the
momentum is more complicated, involving, in the continuum limit, 
several poles and/or cuts. 

This is especially so if a continuous phase transition exists 
in the thermodynamical limit $\Ns\rightarrow \infty$ at 
$\beta_c(\Nt)$.  At the transition
the correlation function is expected to decrease as a power in $z$,
characterized by a critical exponent, the anomalous dimension $\eta$.
As a result, close to $\beta_c(\Nt)$ and for values of $(\Ns,\Nt)$
accessible in practice no simple parametrization of $G(z)$ is
available. Various parametrizations of its Fourier transform in the
continuum limit have been
proposed in the literature (for a review, see Ref. \cite{pelissetto}).
They share the property that, apart from at $\beta_c(\Nt)$, the nearest singularity
remains an isolated pole in the complex plane of the momentum squared.  
Here, for any finite lattice, we will determine the mass gap squared 
as the distance to zero of the nearest pole. 

 We start by transforming the numerical data for the
 correlation function to momentum space. Given
$G(z)$ on the integer values $z= 0,1, ... , \Ns-1$, we 
define $\tilde{G}_q$ via

\be
G(z) = \sum_q e^{-iqz}\tilde{G}_q, \label{fourier}
\ee

\noindent
where  
\be
q = \frac{2\pi n}{\Ns}; \hspace{1cm} n = 0,1, ... , \Ns-1.  \label{qi}
\ee

\noindent
Using the symmetries of the action, the periodic boundary conditions on the 
lattice and the reality of $G(z)$, it is suitable to rename
$\tilde{G}_q$ as $\tilde{G}(p^2)$, where to $q$ we associate
\be
p = 2\sin (\frac{q}{2}).  \label{pi}
\ee

Given numerical data for $G(z)$ we compute the 
susceptibility $\chi$ of the Polyakov loop and the inverse 
of its Fourier transform $H(p^2)$, 
\bea
\tilde{G}(p^2)&\equiv& \frac{1}{\Ns} \sum_{z=0}^{\Ns-1} 
\cos (qz) G(z), \label{gp}\\
\chi &=& \Ns^2 \tilde{G}(0),  \label{chi}\\
H(p^2)&\equiv& \frac{1}{\tilde{G}(p^2)}.  \label{hp2}  
\eea
For these measurements we restrict ourself to the disordered phase,
where the subtraction of $|\av{L}|^2$ is not necessary. On a finite
lattice $\av{L}=0$ also in the ordered phase
in simulations which are long enough because of tunneling
 between the degenerate vacua.

Up to now, the functions $\tilde{G}(p^2)$ and $H(p^2)$ are
defined only on the discrete values
given in Eqs. (\ref{qi}, \ref{pi}). We now extend these functions to 
arbitrary complex values of $p^2$, and define the mass $m$ as
the first zero of $H(p^2)$ on the negative real axis,
\be
H(-m^2) = 0  \label{mgap}.
\ee

In the OZ approximation, we trivially get 
\bea
H_{OZ}(p^2)&=&a_0+a_1\,p^2, \label{hoz} \\
m^2&=&\frac{a_0}{a_1}.   \label{moz} 
\eea
Within our
assumptions, the point $p^2=-m^2$, 
an isolated single pole of $\tilde{G}(p^2)$, is
generically (in the thermodynamical limit) inside 
the circle of convergence of the series expansion of $H(p^2)$ around zero.
Because on the one hand we focus on the long range properties
of the correlation, and on the other hand wish to eliminate as much as possible
discretization effects, we truncate the full series of $H(p^2)$
to a small order $n_{max}$ in $p^2$, and determine  its coefficients
from small $p^2$ data only. We write

\be
H(p^2)=\sum_{n=0}^{n_{\mathrm{max}}} a_n (p^2)^n, \label{hnmax}
\ee
and solve for $a_n$ the system of $n_{max}+1$ equations 
\be
\sum_{n=0}^{n_{\mathrm{max}}} a_n (p_i^2)^n=H(p_i^2) 
\quad;\quad i=0,1, \cdots,n_p  \label{solve}
\ee
where $p_i$ are the discrete values defined by Eqs. (\ref{qi}, \ref{pi})
and the right hand sides are measured  
using Eqs. (\ref{pi},\ref{gp},\ref{hp2}) in $n_p$ points. Then $m^2$
is the smallest positive solution of Eq.(\ref{mgap}).
The case $n_{max}$=1
is the OZ approximation (\ref{hoz}); the results presented
in section 3 in the context of the scaling behaviour of $m$  
are those obtained with $n_{max}$=2 and $n_p=3$. 
The errors quoted there are statistical
errors only, estimated by a bootstrap technique applied to the whole
set of configurations measured.

\medskip

\subsection{The setup for the critical scaling analysis}

\medskip
Near the phase transition one may expect that for given $\Nt$ the
theory is described by an effective $Z_3$ symmetric two-dimensional model for
the Polyakov loop (\ref{polyak}), which is in the same universality
class as the two dimensional three state Potts model \cite{svetitsky}.
It should therefore have the same two independent
critical exponents $\nu$ and $\eta$ as in the 
latter model. These are known from analytic calculations in that model to be 
\bea
\nu & = & 5/6, \label{nu} \\
\eta & = & 4/15. \label{exp}
\eea
The universality hypothesis is supported by the results of  \cite{legeland}. There  
the critical exponents reported for $\Nt=2,4$ and 
$6$ are, within errors, which are around $20\%$ always compatible  
with the above expectations. 

Using the definitions (\ref{gp}-\ref{mgap}), we  
will assume finite size scaling behaviour for $\Ns$ large enough. Thus
the mass gap and the susceptibility are represented
close to the transition by
\bea
\Ns\,m & = & f_1(s), \label{f1}\\
\frac{\chi}{\Ns^{2-\eta}} & \equiv & 
\Ns^\eta\,\tilde{G}(0)=f_2(s), \label{f2}
\eea
where
\be
s = \Ns^{1/\nu}(1-\beta/\beta_c(\Nt)), \label{s}
\ee
and
\bea
f_1(s\to\infty )&\propto& s^\nu.  \label{s1} \\
f_2(s\to\infty )&\propto& s^{-\nu(2-\eta)}.  \label{s2}
\eea
Here, $\beta_c(\Nt)$, yet to be determined, denotes the critical 
value of the gauge coupling in the thermodynamic limit.
The prefactors on the left hand sides of (\ref{f1},\ref{f2})
are such that for $\beta$ close to $\beta_c(\Nt)$ 
the correlation length and the susceptibility behave in the 
$\Ns\to\infty$ limit as 
prescribed by the critical exponents, that is
\bea
m&\propto& (1-\beta/\beta_c(\Nt))^{\nu}, \\
\chi&\propto& (1-\beta/\beta_c(\Nt))^{-\nu(2-\eta)}.
\eea

In the following analysis, we assume universality to be true and use
the values of $\nu$ and $\eta$ given in 
Eqs.(\ref{nu}, \ref{exp}). A posteriori the fact that the data fall on
unique scaling curves show that these exponents are the correct ones for
the $SU(3)$ gauge theory in two spatial dimensions.
For the susceptibility we make a further test of
this assumption showing that the data do not fall on a universal function $f_2(s)$
if we assume the mean field exponent $\eta=0$.
    
In the next section, we use finite size scaling applied to a different 
quantity, more easily  measurable on the lattice, in order to 
determine $\beta_c(\Nt)$ for
$\Nt=4,6,8,12$. Then the scaling assumptions of Eqs.(\ref{f1} - \ref{s2}) 
are checked with a high accuracy in the case $\Nt=6$, 
for which a large density of data points in parameter space has
been acquired. Finally an extrapolation of $\beta_c(\Nt)$ 
to the continuum limit $\Nt\to\infty$ is performed, which allows to
compare the gauge theory and the string model near their transition points.

\section{Applications of finite size scaling. Critical coupling, universal
scaling functions and the
continuum limit compared with the string model}

\subsection{The critical couplings $\beta_c(\Nt)$ }

We recall here the properties of finite size scaling, which we need to
determine the critical coupling.
We determine $\beta_c(\Nt)$ from the condition
that the average value $\Phi_L$ of a  {\it {classically dimensionless}} 
functional $\Phi\{L\}$ of the effective field $L(\bar x)$ defined by 
Eq.(\ref{polyak})
does possess finite size scaling properties analogous to those of $m$ and $\chi$
in Eqs. (\ref{f1},\ref{f2}). Generically, we set
\bea
\Ns^{\phi}\,\Phi_L  & = & f(s) \label{phi} \\
s & = & \Ns^{1/\nu}(1-\beta/\beta_c(\Nt)) \nonumber \\
f(s\to\infty )&\propto& s^\rho,  \label{f}
\eea
where $\phi,\nu,\rho$ are critical exponents.  The conditions that,
in a domain including $\beta = \beta_c(\Nt)$, 
 
 i) $\Phi_L$ is well defined on the lattice for any $\Ns$
 
 ii) $\Phi_L$ exists in the thermodynamic limit 

\noindent
imply the relation
\be
\rho=\nu\,\phi. 
\ee
In this limit
Eq.(\ref{phi}) then gives
\be
\Phi_L\,\propto\,(1-\beta/\beta_c(\Nt))^{\nu\phi},
\ee
and $\phi$ is the anomalous dimension of $\Phi$. 
If $\beta$ approaches $\beta_c(\Nt)$, one may
expand $f(s)$ around zero  and rewrite (\ref{phi}) as
\be
\Ns^{\phi}\,\Phi_L =a + b\Ns^{1/\nu}(1-\beta/\beta_c(\Nt)) + ...,
\label{expand}
\ee
where $a$ and $b$ are constants.
The right hand side is thus linear in $N_S^{1/\nu}$ at $\beta$ fixed,
and $\beta_c(\Nt)$ is that value of $\beta$ for which the
slope $b(1-\beta/\beta_c(\Nt))$ vanishes.
The method is, of course, particularly
useful if $\phi$ and $\nu$ are known. 

\medskip

The order parameter $\av{L(\bar{x})}$, which is classically dimensionless
and has an anomalous dimension $\eta/2$ in two dimensions, does not fulfill the condition
i) above since, on the lattice, it vanishes even in the broken phase,
due to fluctuations between degenerate groundstates. For the same reason
the susceptibility, which is quadratic in $L$ and
thus has the anomalous dimension $\eta$ (see Eqs.(\ref{f2},\ref{s2})),
is not suitable because its measurement via the definition 
(\ref{chi}) requires
$\tilde{G}(0)$ to be the {\it {connected}} correlation (by subtracting
$|\av{L}|^2$). One could in principle use $\tilde{G}(0)$ without subtraction, but we have 
found that it is less stable over the transition than the variable defined below.

\begin{figure}[!ht]
\begin{center}
\includegraphics[width=\figwidth]{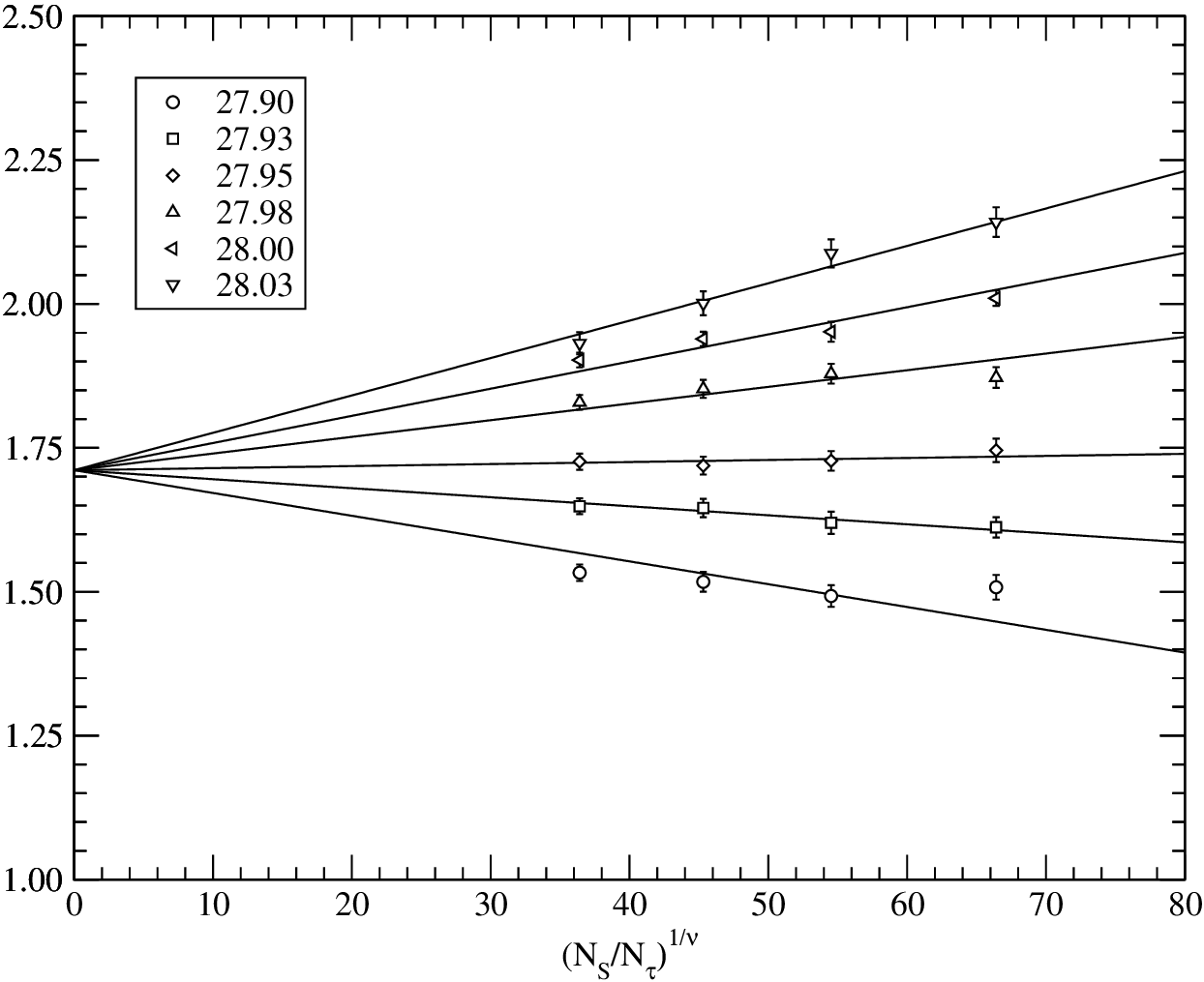}
\end{center}
\vspace{-0.8cm}
\caption{\label{fig:L3_06} $N_S^{3\eta/2}{\cal L}_3$ as a function of $\Ns^{1/\nu}$
for fixed values of $\beta$. The temporal extent is $\Nt=8$. The straight lines
correspond to the fits described in the text.}
\end{figure}

In fact, we consider the quantity ${\cal L}_3$ defined as 
follows
\bea
{\cal L}=\frac{1}{\Ns^2}\sum_{\bar {x}}L(\bar x),  \label{calL} \\
{\cal L}_3\equiv  Re \av{{\cal L}^3}.     \label{calL3}
\eea

\noindent
This quantity is particularly interesting. $\cal L$ is the lattice average
of $L(\bar{x})$, and its cube is the monomial of lowest degree that does
not suffer from the same disease as  $\av{L}$ itself:  
in the ordered phase ${\cal L}^3$ fluctuates around the same real 
value whatever vacuum is chosen during the simulation.  We have checked
that the imaginary part of $\av{{\cal L}^3}$ is always negligible, due to
the reality of the action, and use its real part in
(\ref{calL3}) for convenience.

\begin{figure}[!ht]
\begin{center}
\includegraphics[width=\figwidth]{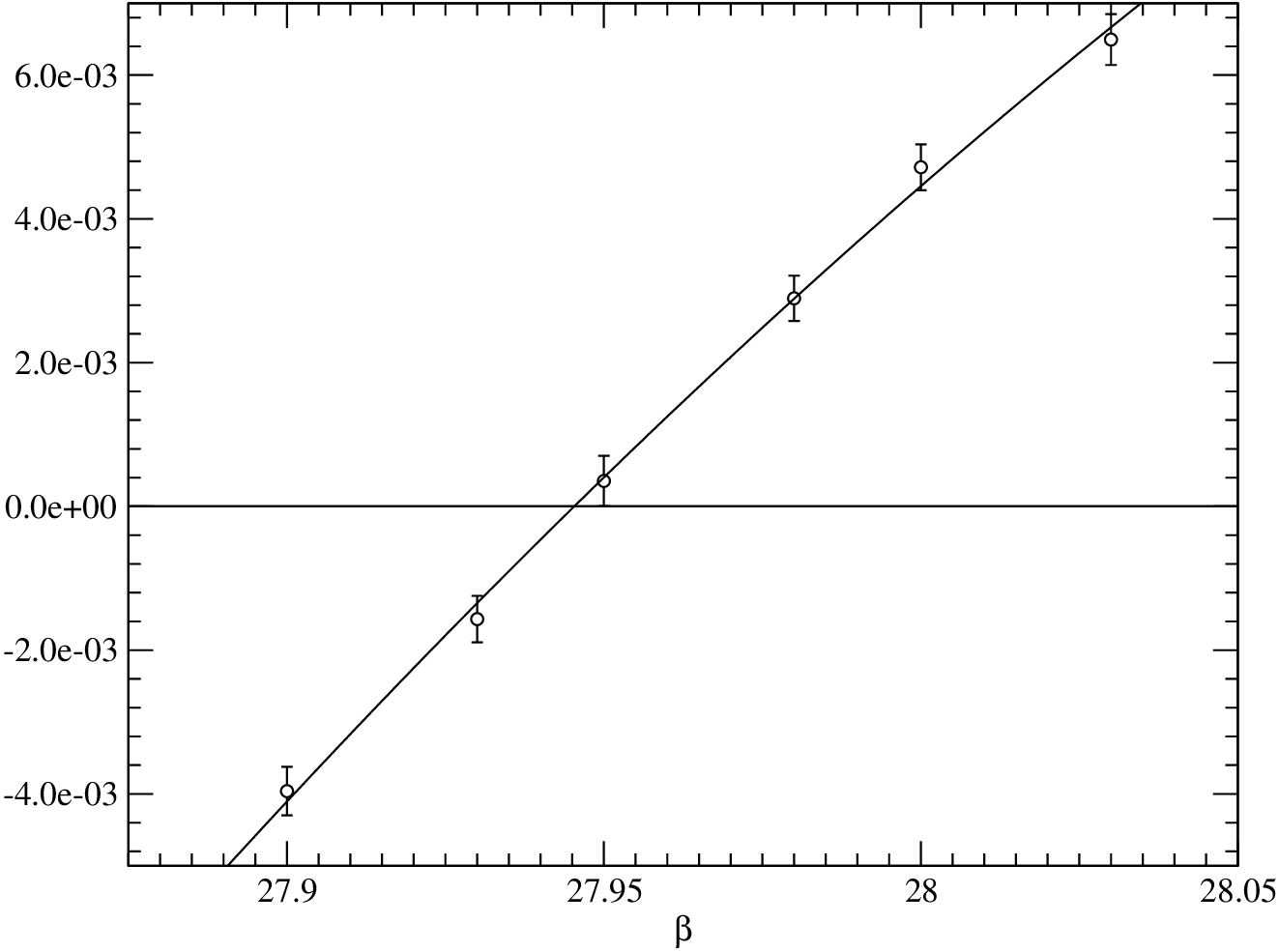}
\end{center}
\caption{\label{fig:slopes_L3_08} The slopes
 $\xi(\beta)$ of $\Ns^{3\eta/2}{\cal L}_3$ for $\Nt=8$
as a function of $\beta$. The curve corresponds to the fit described in the text.}
\end{figure}

In Fig.\ref{fig:L3_06} we show  data obtained for ${\cal L}_3$
at $\Nt=8$ inside a domain in $\beta$ which strongly suggests
that it contains a value of $\beta$ such that
${\cal L}_3$ is independent of $\Ns$. More precisely, inside
the domain in $\{\beta,\Ns\}$ shown, it appears that the data
are consistent with ${\cal L}_3$ being linear in 
$\Ns^{1/\nu}$, as expected if  an expansion of the type of Eq. (\ref{expand})
is valid: The straight lines drawn actually result from linear fits 
to the data at $\beta$ fixed. They are restricted to go through
the same point for vanishing argument, as demanded
by Eq. (\ref{expand}). Similar results are obtained for the other values
of $\Nt$.
In addition, the same ansatz (\ref{expand}) implies that the slopes 
of ${\cal L}_3$ in $\Ns^{1/\nu}$ at $\Nt$ fixed are proportional
to $1-\beta/\beta_c(\Nt)$ near $\beta_c$. To take care of higher order
corrections we fit the slopes, which we denote by $\xi(\beta)$, using

\be
\xi(\beta)=a_1(\beta-\beta_c)+a_2(\beta-\beta_c)^2,
\ee
where $a_1$ and $a_2$ are the parameters of the fit.
These fits have very good $\chi^2/d.o.f.\simeq 1$. In
Fig.\ref{fig:slopes_L3_08} we show as an example the fit for $\Nt=8$,
which confirms the nearly linear behaviour of the slope near $\beta_c$. 
The same is true for the other values of $\Nt$
exploited.
Those fits then
complete the determination of  $\beta_c(\Nt)$ shown in the Table below. Note that
the errors are the statistical ones only. The critical coupling $\beta_c(4)$ has been
determined in two earlier works \cite{legeland} \cite{liddle}. The values were
$14.74(5)$ and $14.717(17)$ respectively. The method in these investigations are different
from each other and from ours. Systematic errors have not been taken into account.
Therefore the results are in very good agreement with ours, giving further support 
to our method. In Ref. \cite{smekal} still another method is used to estimate the
critical coupling in this theory for $\Nt=4,6,8$. For $\Nt=6,8$ the values are in agreement
with ours within their errors, which are ten times larger. For $\Nt=4$ there is a discrepancy,
but this may be due to the fact that their procedure has not converged on the spatial
volumes they use, which are considerably smaller than ours.

\begin{center}
\begin{table}[!ht]
\begin{center}
\begin{tabular}{|c|c|}
\hline
$\Nt$ & $\beta_c(\Nt)$ \\
\hline
\hline
$4$      & $14.7404 (7)$ \\
$6$      & $21.374  (3)$ \\
$8$      & $27.952  (4)$ \\
$12$     & $41.075  (7)$ \\
\hline
\end{tabular} \label{tab:beta_table}
\end{center}
\end{table}
\end{center}

\subsection{Finite size scaling for $m$ and $\chi$}

Knowing $\beta_c(\Nt)$, we are now able to test the universal scaling of 
the mass gap and the susceptibility. They are defined in
Eqs.(\ref{mgap},\ref{chi}). 
Their expected finite size scaling behaviours were described in 
Eqs.(\ref{f1} - \ref{s2}). 
Here we illustrate the fact that,
apart from a known $\Ns$ dependent prefactor, they depend on the two parameters 
$\beta$ and $\Ns$  via a function of the single variable $s$ of Eq.
(\ref{s}). We have measured the correlations, and thus $m$ and $\chi$
for $\Nt=4$ and $6$. Here we concentrate on $\Nt=6$, our conclusions 
for  $\Nt=4$ being identical. We have used values for $\Ns$ from $48$
to $264$, and $20$ different $\beta$-values between $\beta=18$ and
$21.35$, just below $\beta_c=21.374$.

\begin{figure}[!ht]
\begin{center}
\includegraphics[width=\figwidth]{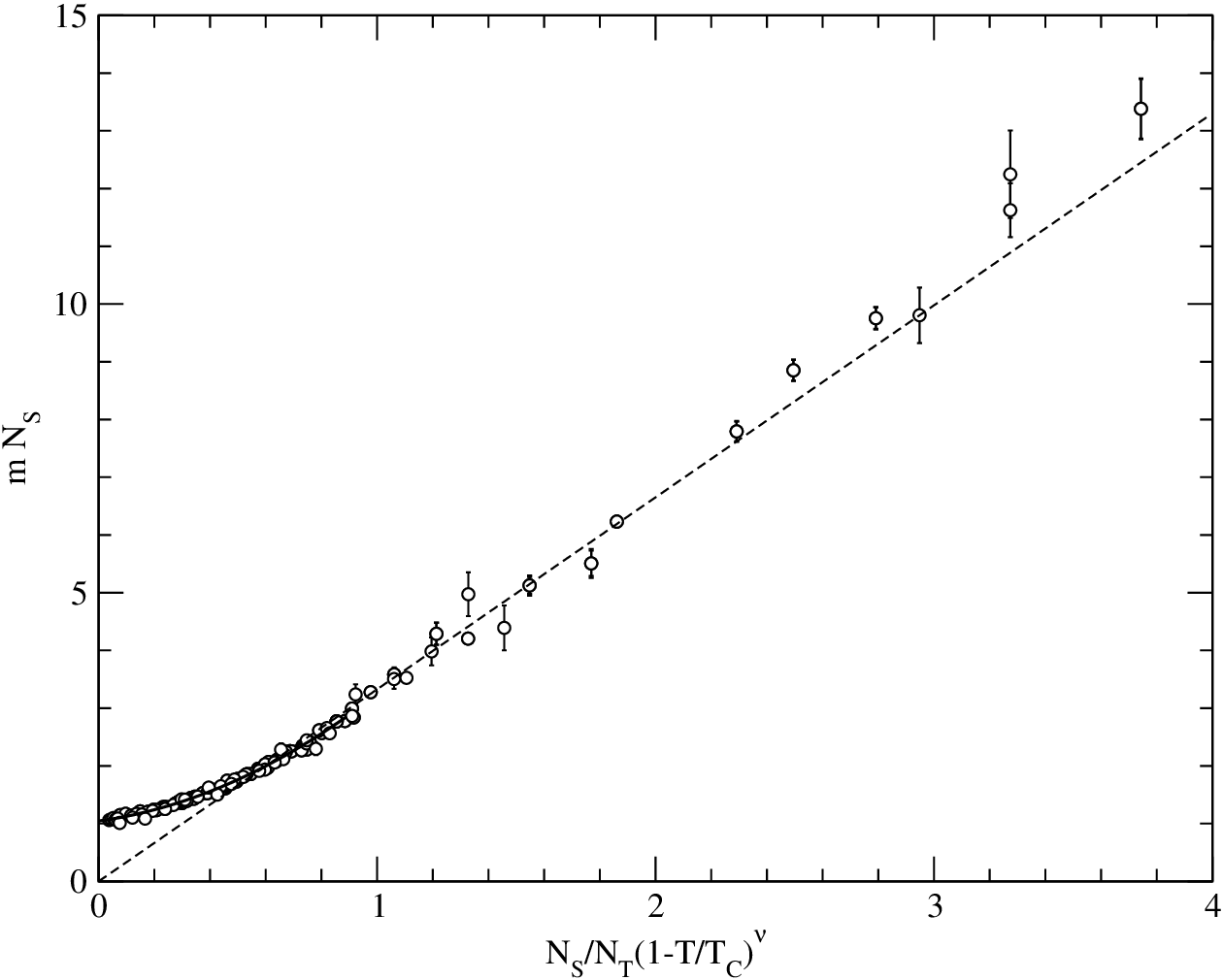}
\end{center}
\caption{\label{fig:Fig3.m3.eps} The scaling behaviour of the mass gap
at $\Nt$=6. The full line is the finite size scaling parametrization discussed in the text.
The dashed line is the asymptotic behaviour given in Eqs. (\ref{s},\ref{s1}).
}
\end{figure}

We start with the mass because in this case only the critical 
exponent $\nu$ enters the analysis and thus can be probed independently. 
The scaling properties of the mass described by
Eqs.(\ref{f1},\ref{s},\ref{s1}) are illustrated for $\Nt=6$
in Fig. \ref{fig:Fig3.m3.eps}. The quantity $m\Ns$ is plotted 
versus $t^{\nu}/\Nt$,
where

\be
t=\Ns^{1/\nu}( 1-T/T_c). \label{t}
\ee
The temperature $T$ is determined through its definition (\ref{temp}),
with the lattice spacing $a(\beta)$ obtained from a fit to the
zero temperature string tension, given in our earlier paper \cite{bialas4},
and also repeated below in Eqs. (\ref{sig0} - \ref{p}). It is clear from
the definition in (\ref{t}) that $t\simeq s$ near the transition, and thus can be used as
a finite size scaling variable.
The factor $1/\Nt$ is inserted because
according to (\ref{msigma}), the continuum limit of the product $m\Nt$ depends
only on the temperature, and thus provides the leading term $(1-T/T_c)^{\nu}$
as $\Nt \to \infty$. The values of $m\Nt$ collected in our earlier paper
\cite{bialas4} 
for $\Nt=4$ and $6$ are, in fact, found to be close one to the other.
 
Up to $t^{\nu}/\Nt\simeq 1$
($t\simeq 9$) 
, the existence of a
finite size scaling law is impressively 
confirmed, showing a high density of points lying on one single curve.
The curve is a fit to a second degree polynomial in $t$.
This validates the value $\nu=5/6$  expected from universality. For
$t^{\nu}/\Nt\geq 1$
 the data in the figure lie in the neighborhood of a straight line,
corresponding to the asymptotic behaviour of Eq. (\ref{s1})
with the same value $\nu=5/6$.

\begin{figure}[!ht]
\begin{center}
\includegraphics[width=\figwidth]{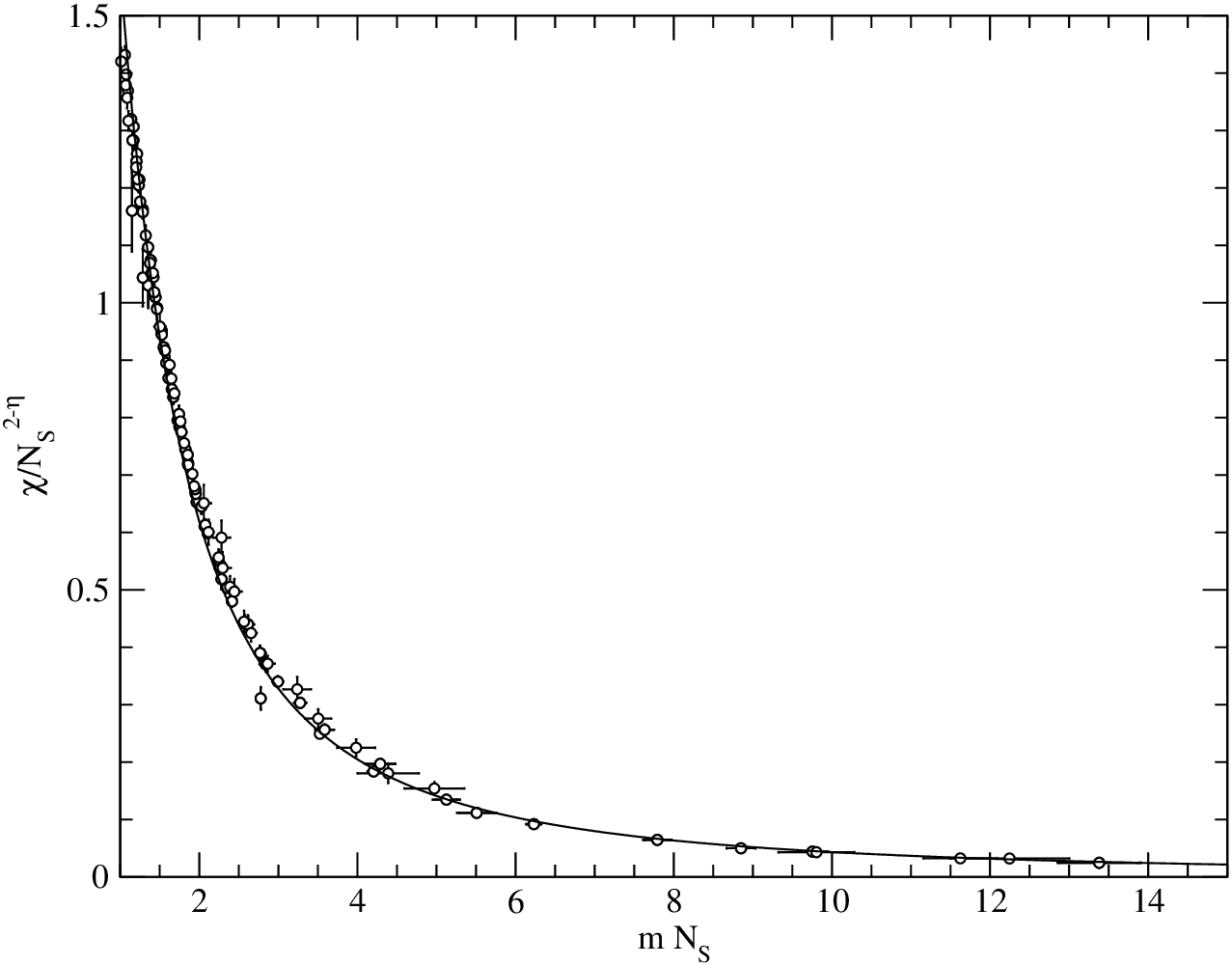}
\end{center}
\caption{\label{fig:article_ns_g0_color_06.eps} The scaling behaviour of
the susceptibility at $\Nt$=6. The curve corresponds to the fit described
in the text.
}
\end{figure}

We now probe the anomalous dimension $\eta=4/15$ appearing in
the left hand side of Eq.(\ref{f2}). This is done in Fig.
\ref{fig:article_ns_g0_color_06.eps} by plotting $\chi/\Ns^{(2-\eta)}$ against
$m\Ns$. This choice eliminates the explicit dependence on the critical
exponent $\nu$, and therefore is a direct test of $\eta$. As we can see in
the figure all the data fall on a unique scaling curve.
This successfully validates $\eta=4/15$. The curve in the figure
corresponds to a fit
\be
\frac{\chi}{\Ns^{2-\eta}} = \frac{b_1}{1+(b_2m\Ns)^{(2-\eta)}} \label{chifit}
\ee
where $b_1=4.77(5)$ and $b_2=1.50(1)$. The form of the function has been chosen to
be consistent with the expected large $\Ns$ behavior. This behavior can be easily derived
from Eqs. (\ref{f1} - \ref{s2}).

As already mentioned, to measure the mass gap and
susceptibility above $\beta_c$ ($s<0$) one should 
subtract from $G(z)$ the disconnected contribution  $|\av{L}|^2$. 
On a finite lattice $\av{L}$ vanishes,
because of the tunneling between the degenarate vacua. In practise
this happens in Monte Carlo simulations near the critical point 
for $\Ns$ not too large. 
A popular way out is to replace $|\av{L}|^2$ by $\av{|L|^2}$. 
We will not use this procedure here. By the way, our
choice to use ${\cal L}_3$ to identify the critical $\beta$ was
dictated by the need for a control parameter which evolves smoothly
across the transition.


\begin{figure}[!ht]
\begin{center}
\includegraphics[width=\figwidth]{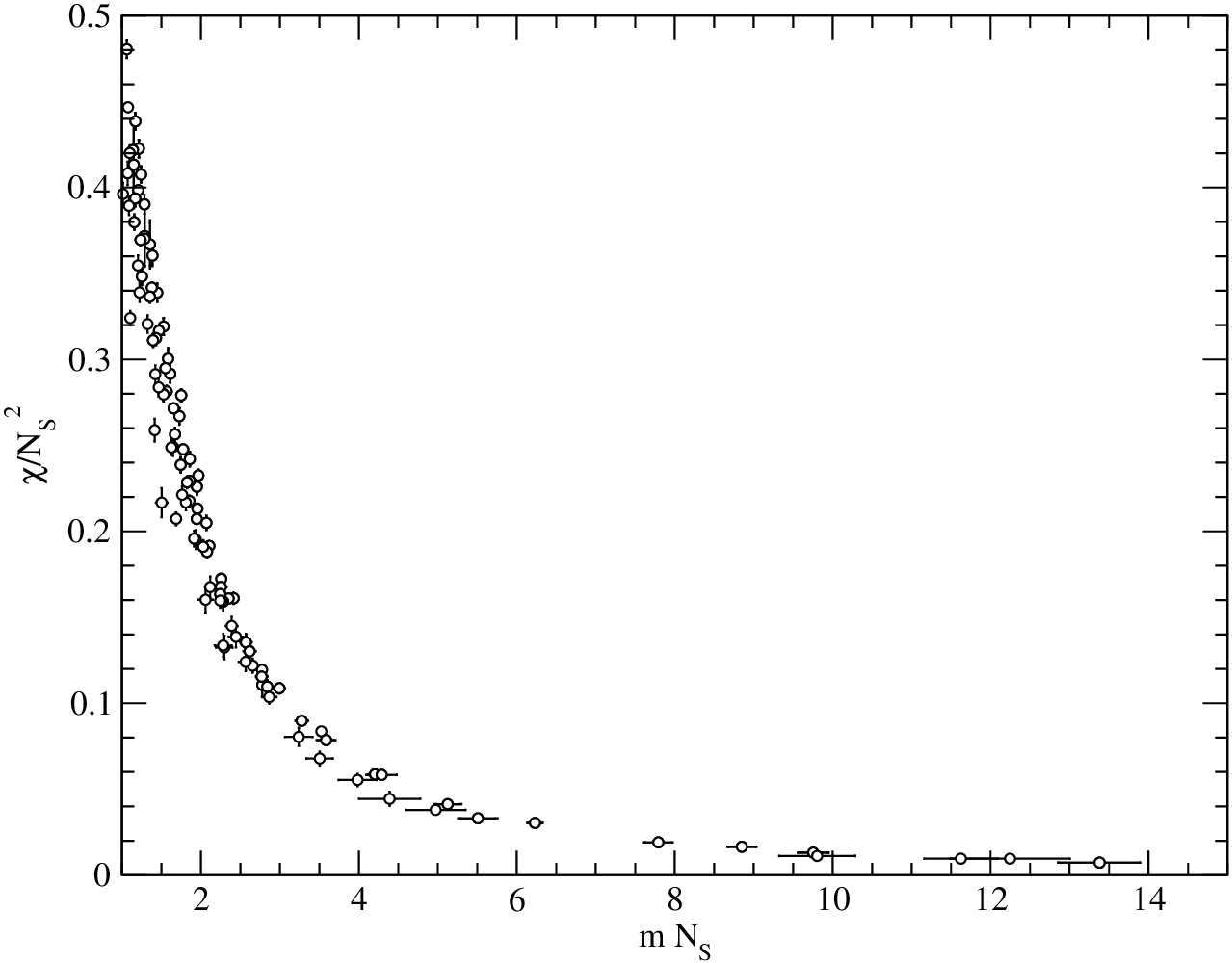}
\end{center}
\caption{\label{fig:article_ns0_g0_color_06.eps}  This figure shows 
the absence of a unique scaling function for the assumption of mean
field exponents.
}
\end{figure}

The analysis described above, which has also been performed for $\Nt=4$
with the same conclusions convincingly comforts the expectation that the
critical behaviour of the gauge model
is governed by the same exponents as 
the two-dimensional  three states Potts model. We have not tried to determine
these exponents a priori from demanding a unique scaling curve for 
each of the variables considered, ${\cal L}_3,m,\chi$. 
But, taking the susceptibility as an example, we show a contrario  
in Fig. \ref{fig:article_ns0_g0_color_06.eps} that
choosing the mean field exponent
$\eta=0$ is far from giving  
a unique scaling curve. 

\subsection{The critical temperature in the continuum compared with
  the string model}

As shown above, for fixed $\Nt$ there is a phase transition at a critical
coupling $\beta_c(\Nt)$ in the
thermodynamic limit $\Ns\rightarrow\infty$. In the continuum theory
there is thus a phase transition at the critical temperature
$T_c$, given 
from Eqs. (\ref{temp}, \ref{g2}) by
\be
\frac{T_c}{g^2}=\lim_{\Nt\rightarrow\infty}\frac{\beta_c(\Nt)}{6\Nt}
\ee 
To determine this ratio we use the data given in Table 1, and assume
that the approach to the continuum limit is given by
\be
\frac{\beta_c(\Nt)}{6\Nt}=
\frac{T_c}{g^2} (1+ \frac{b}{\Nt}+\frac{c}{\Nt^2}). \label{tg2fit}
\ee
>From the data for the largest values of $\Nt$, it is clear that the leading
correction is linear in $1/\Nt$. Inserting the values from Table 1 
in Eq.(\ref{tg2fit}), we obtain a quite good fit.
We obtain 

\be
\frac{T_c}{g^2}=0.5446(4).  \label{tcg2}
\ee

\medskip Another choice is to use the square root of the zero
temperature string tension as energy scale. This is also what we need
to compare our results with the string model.  In our earlier paper
\cite{bialas4}, we used data in the literature \cite{liddle}
\cite{teper2} to obtain the following parametrization: \be
a\sqrt{\sigma_0}\equiv F_{\sigma_0}(\beta) =\frac{h}{\beta}\frac{\beta - z}{\beta
  - p} \label{sig0}
 \ee 
where
 \bea
h & = & 3.325(6)\\
z & = & 2.0(2)\\
p & = & 3.7(1) \label{p}
\eea
The form of the fitting function was choosen from the observation that $\beta F_{\sigma_0}(\beta)$
has a finite value when $\beta\rightarrow \infty$, and is a slowly varying function of $\beta$
in the region of the data to be fitted. A posteriori we find that the fit is very good and the
value of the zero and the pole  are far from the region of the data used.
 
On should be aware of the fact that the errors on the parameters are strongly
correlated. What is important is the error on the interpolating
function $F_{\sigma_0}(\beta)$.  We found that the absolute value of
the error on $1/F_{\sigma_0}(\beta)$ grows from 0.002 for $\beta=14$
to 0.01 for $\beta=42$. The effect of this error on $T_c/\sqrt{\sigma_0}$
was estimated using the bootstrap technique.

To obtain  the continuum value of 
$\sqrt{\sigma_0}$ normalized to the scale setting parameter
$g^2$, we form the ratio
\be
\frac{\sqrt{\sigma_0}}{g^2}=\lim_{\beta\rightarrow\infty} 
\frac{\beta F_{\sigma_0}(\beta)}{6}=\frac{h}{6}=0.554(1)\label{sigmag2}
\ee
This value is in agreement with an earlier investigation which gives a value
$0.5528(8)$ for the same quantity \cite{bringoltz2}, validating our parametrization
of the data. 

\medskip
The ratio $T_c/\sqrt{\sigma_0}$ in the continuum limit is obtained from the ansatz
\be
\frac{1}{\Nt F_{\sigma_0}(\beta_c(\Nt))} =
\frac{T_c}{\sqrt{\sigma_0}}(1 + \frac{c_1}{\Nt^2} + \frac{c_2}{\Nt^4}).  \label{tcsigmafit}
\ee
We find
\be
\frac{T_c}{\sqrt{\sigma_0}} = 0.986(1) \label{tcsigma}
\ee
with $\chi^2/d.o.f=0.07$. This is lower than the value 
$T_c/\sqrt{\sigma_0} = 0.999(4)$ in Ref. \cite{liddle}, where the same extrapolation
formula is used, but for smaller values of $\Nt=2-5$.

\begin{figure}[!ht]
\begin{center}
\includegraphics[width=\figwidth]{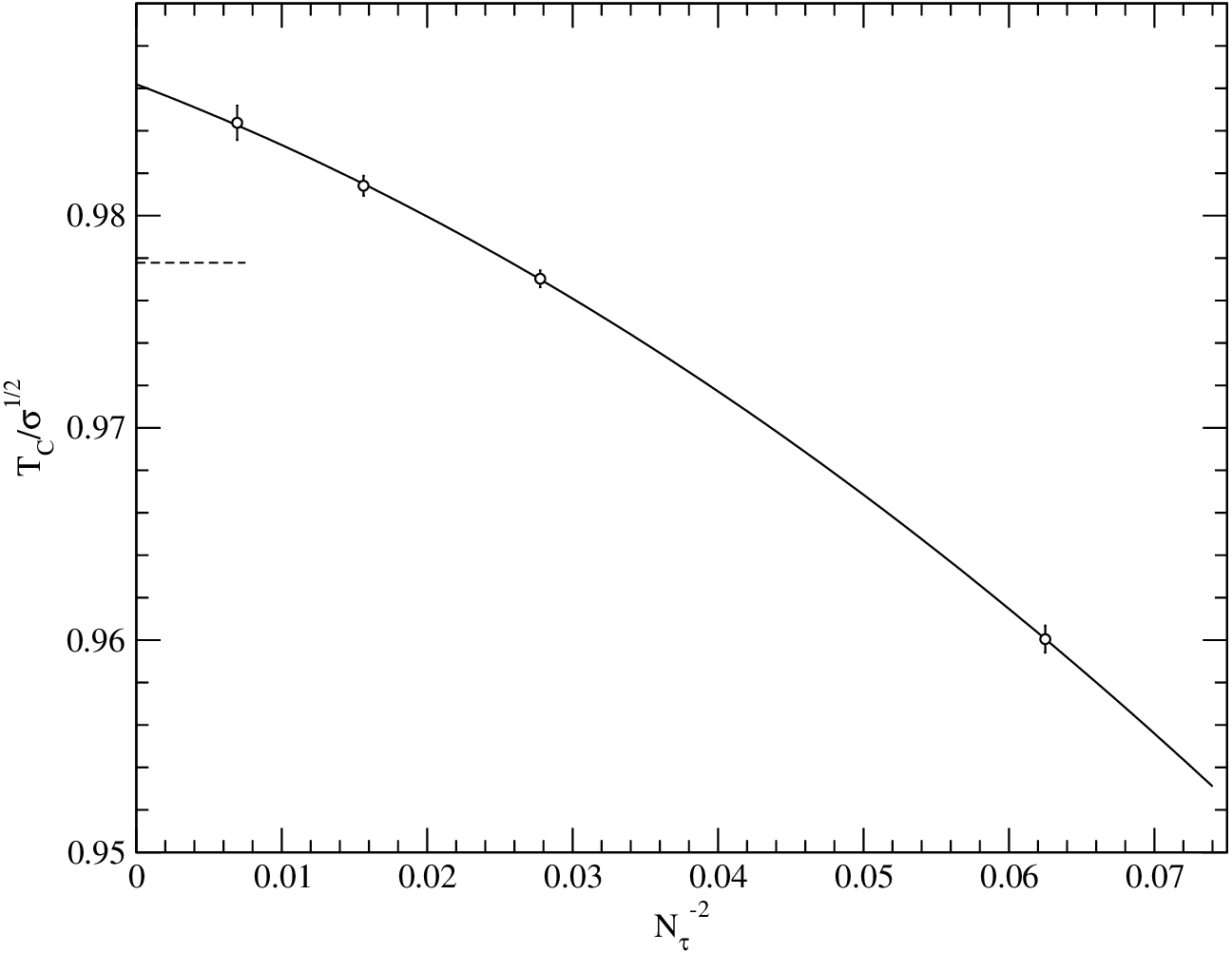}
\end{center}
\caption{ The quantity $T_c/\sqrt{\sigma_0}$ as a function of $1/\Nt^2$. The full line
corresponds to the fit by Eq. (\ref{tcsigmafit}). The dashed line is the string model value in Eq.
(\ref{tcstring}).}
\end{figure}

As can be seen in Fig. 6
 the data show
clearly that the leading correction to
the continuum value in (\ref{tcsigmafit}) has to be quadratic in $1/\Nt$ as
expected for a ratio between physical quantities.

In the Nambu-Goto string model, the temperature dependent string
tension is predicted to be \cite{olesen} 
\be 
\sigma(T)=
\sigma_0\sqrt{1-\frac{T^2}{T_c^2}} \label{string1} 
\ee 
where \be
T_c^2=\frac{3\sigma_0}{(D-2)\pi} \label{string2}
\ee 
Here the constant $D$ is the
number of space time dimensions of the gauge model. The prediction
only depends on the number of transverse dimensions and not on the
group $SU(N)$. In our case which has one transverse dimension it leads
to 
\be 
\frac{T_c}{\sqrt{\sigma_0}}= \sqrt{\frac{3}{\pi}}=0.977... \label{tcstring} 
\ee 
Furthermore Eq. (\ref{string1}) shows
that the approach to $T_c$ corresponds to the mean field critical
exponent $\nu=1/2$.

It is quite astonishing that the measured value (\ref{tcsigma})
for $SU(3)$ is so close
to the theoretical one (\ref{tcstring}), in particular as we have shown that
there is a scaling region, where the mean field behaviour of the string 
model is not valid. In principle it would be
interesting to investigate if the prediction (\ref{tcstring}) becomes better
in $SU(N)$ for larger values of $N$. It has, however, been shown that for $N\geq 5$
the transition is of the first order \cite{holland, liddle2}
as is also predicted from the universality arguments \cite{svetitsky}. 
Then $\sigma(T_c)$ has a jump, and Eq. (\ref{tcstring}) is not expected
to be valid. Values of
$T_c/\sqrt{\sigma_0}$ have also been obtained for $SU(2)$ and $Z_2$ gauge
theories. In these cases the values of this ratio is quite far from the string
value in Eq. (\ref{tcstring}), being $1.12(1)$ \cite{liddle2} and $1.237(3)$
\cite{caselle,hasenbusch} respectively.

\begin{figure}[!ht]
\begin{center}
\includegraphics[width=\figwidth]{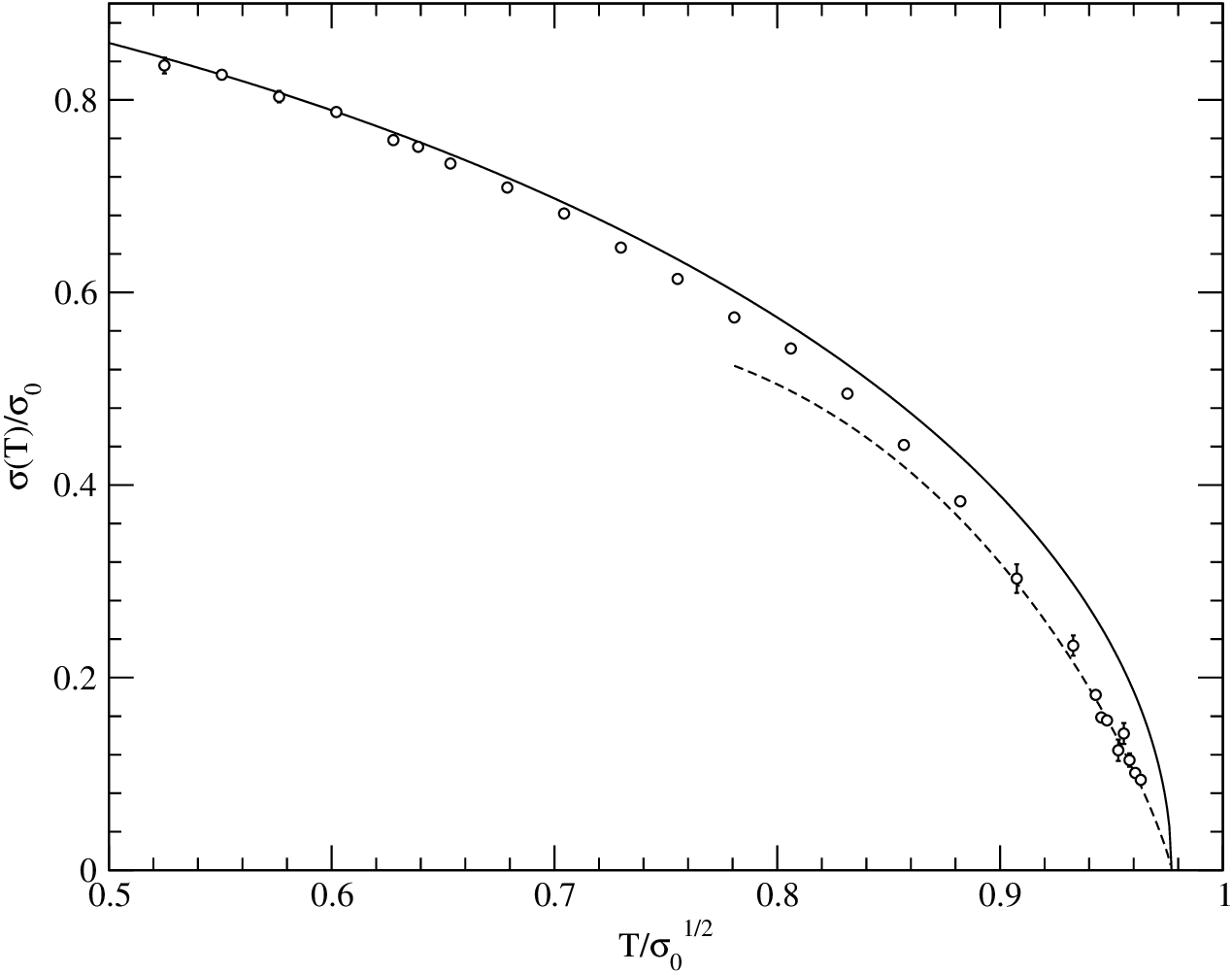}
\end{center}
\caption{\label{fig7}The quantity $\sigma (T)/\sigma_0$ as a function of
$T/\sqrt{\sigma_0}$. The full line corresponds to the string model as described in
the text. The dashed line shows the scaling behaviour near the transition.}
\end{figure}

Finally, we use the analysis of the correlation functions
of the Polyakov loops in the
region $18\leq \beta \leq 21.3$ performed in this paper
 to extend the comparison with the string
model in \cite{bialas4} to the whole range below the transition.
In Fig. 7 we have plotted $\sigma(T)/\sigma_0\equiv m/(\Nt(F_{\sigma_0}(\beta))^2)$
vs $T/\sqrt{\sigma_0}\equiv 1/(\Nt F_{\sigma_0}(\beta))$. In the region near the
transition, we only use the largest volume
for each $\beta$, and further restrict to those cases
where $m\Ns>2.7$ in which case the corresponding points lie along the dashed line in Fig. 3.
 We compare with the
prediction of the string model given in Eqs. (\ref{string1},\ref{string2}),
which does not have any free parameters. The agreement is good up to 
$T/\sqrt{\sigma_0}\approx 0.7$. Near the phase transition we compare instead
with the scaling behaviour corresponding to $m=c(1-T/T_c)^{\nu}$
, where $c$ is a free parameter. This description
is satisfactory down to $T_c/\sqrt{\sigma_0}\approx 0.9$. If the
data in the region in between
can be described by some correction to the simple string picture 
as discussed in \cite{billo,makeenko} is still an open
question. Any description of data like those in Fig. 7 must, however, be aware of the scaling 
region near the phase transition.

\section{Conclusions}

In this article we have presented results for the behaviour of
three dimensional $SU(3)$ gauge theory near the finite temperature
phase transition. 
We have introduced a new and powerful application of finite size scaling
to extract the value of the critical coupling. For this purpose, we have used
the third power of the Polyakov loop,
which evolves smoothly over the transition. To extract $\beta_c$ we assume the critical
exponents of the two dimensional three state Potts model, which are expected to be
valid because of universality arguments \cite{svetitsky}.  
Employing this method, we have determined the critical coupling
on considerably larger lattices both in space and temperature direction than
in earlier investigations. 

That the phase transition is in the universality class of the abovementioned Potts model
is strongly supported by our finite size scaling analysis of the mass gap and the
susceptibility.
Their values were extracted from an analysis
of the correlation function of the Polyakov loop in Fourier space.
We used our measured critical couplings
and these correlation functions at
a large set of values of the spatial extension
and the coupling constant near the transition. 
In this region all the data for the mass gap fall within errors on a unique
scaling curve, if we assume the critical exponent $\nu$ to be that
of the Potts model. This test is sensitive to the exponent $\nu$ only. 
A further check of the universality, which is 
sensitive to the exponent $\eta$ was performed for
the susceptibility in the disordered phase. Also in this case the data
fall on a single scaling curve when we use the expected value
$\eta=4/15$. This is not the case if we e.g. assume the mean field value $\eta=0$.
This analysis was
performed for $\Nt=4$ and $6$ with consistent results.

Finally we have used our values for the critical couplings for 
$\Nt=4,6,8$ and $12$ and the zero temperature string tension determined at
these values from an interpolation formula of data in the literature 
to make an extrapolation to the continuum and obtain the
ratio of the critical temperature to the square root of the zero temperature string
tension. We have compared this ratio to the corresponding result in the
Nambu-Goto string model.
Although the string model predicts a temperature dependent string tension
which has mean field behaviour
in the scaling region, the critical temperature in this model is, in fact, 
very close to the lattice result. The ratio is a non perturbative and
non universal observable.
Thus it is a very important result,
which may imply that the string model describes the spectrum and the multiplicities
of the excited states of the $SU(3)$ gauge model in three space-time dimensions.

\section{Acknowledgments}

The simulations were done on the SHIVA computing cluster at Faculty of
Physics, Astronomy and Applied Computer Science Jagiellonian
University and also on ZEUS CPU and GPU cluster at Academic Computing
Centre CYFRONET in Krakow and on GPU cluster at Bielefeld
University. P.B. is indebted to Edwin Laerman, Olaf Kaczmarek and
Marcus Fischer for the possibility to use their code and the machine
and for their usual hospitality during his stay at Bielefeld. B.P. is
very grateful for the kind hospitality of the Institut de Physique
Th\'eorique de Saclay, where part of this work was done.

\end{document}